\documentstyle[12pt,psfig,rotate]{article}
\begin{document}
\title{Photoproduction of pions and etas in nuclei\thanks{Work supported by BMBF and GSI Darmstadt}}
\author{M. Effenberger, A. Hombach, S. Teis and U. Mosel\\
Institut f\"ur Theoretische Physik, Universit\"at Giessen\\
Heinrich-Buff-Ring 16, D-35392 Giessen\\
UGI-96-22}
\date{}
\maketitle
\begin{abstract}
We calculate the cross sections for inclusive one-pion, two-pion
and eta photoproduction in nuclei
in the photon energy range from 300 MeV to 900 MeV 
within the framework of
a semi-classical BUU transport model.
Our results are compared with existing experimental data and discussed
with respect to a calculation of the total photoabsorption cross
section.
\end{abstract}
\section{Introduction}
In a recent work \cite{abspaper} we have performed a calculation of the total
photoabsorption cross section on nuclei within a
semi-classical phase space model in order to understand the experimentally
observed medium modifications compared to the elementary cross section on the
proton and deuteron: 
the broadening of the $\Delta$-peak and the
disappearance of the higher resonances $D_{13}$ and $F_{15}$
\cite{frommhold,bianchi,daphne96,armstrongproton,armstrongdeut}. 
We calculated
the in-medium widths of the involved nucleon resonances by using estimates for
the cross sections for resonance-nucleon scattering. It turned out
that the collision widths, especially for the $D_{13}$, were much too small
to explain the mentioned medium modifications; a collision width of 300
MeV that has been ascribed to the $D_{13}$-resonance by phenomenological
fits to the absorption cross section \cite{alberico,kondr94,bianchi2} seems far
too large. In the $\Delta$-region the missing strength in the
excitation function indicates that many-body absorption mechanisms should
be included.
\par We now want to look at more exclusive reaction channels in order to gain
further insight into the photon-nucleus reaction
and have therefore calculated cross sections for one-pion, two-pion and
eta production in nuclei. 
In these studies we have used a BUU transport model that employs the very
same microscopic ingredients as the calculations in \cite{abspaper} and
has been
very successfully applied to the description of
heavy-ion collisions up to bombarding energies of 2 GeV/A \cite{ber84,cass90,teis,wolf} and
pion-nucleus reactions \cite{engel93}. 
We thus expect that the incoherent interaction of the primary produced mesons and nucleon
resonances with the surrounding nuclear medium is quite well simulated.
The non-perturbative treatment of all processes following the photon
absorption makes the BUU model an excellent tool for these calculations.
Therefore it should be possible to trace back medium modifications of
the elementary photon-nucleon interaction.
\par Presently the total cross section for photoproduction of pions in nuclei
has only been measured up to photon
energies of 500 MeV \cite{arends}. Theoretical calculations within the
framework of the $\Delta$-hole model \cite{carras92} and the 
coupled channel BUU model
\cite{hombach} were able to explain the size of the observed cross sections
but failed in the description of the broad structure in the $\Delta$-region.
\par Pion photoproduction on nuclei is determined both by the elementary 
$(\gamma,\pi)$ process on the nucleon as well as by final state $\pi$-N
interactions whereas photoabsorption is dominated by the former reaction.
A detailed investigation of $(\gamma,\pi)$ on nuclei could thus help to
separate these two effects and to identify true in-medium effects on the
primary production process.
A possible medium modification might, for example, be a strong modification of the
elementary two-pion production cross section or a lowering of the
$\rho$-meson mass in nuclei \cite{hatsuda} that would affect the width of the
$N(1520)$-resonance.
\par In section \ref{model} we start with a brief presentation of the used
BUU model. The parameterizations of the elementary cross sections follow
in section \ref{ele} which are used to calculate the photoproduction of
pions (section \ref{pionprod}) and etas (section \ref{etaprod}).
\section{The BUU model}
\label{model}
Since the BUU model has been described in full detail in the
literature (e. g. \cite{ber84,cass90}) we restrict ourselves
here to the basic ideas and focus on the points that
are of particular interest for the calculations presented here.
The BUU equation describes the classical time evolution of a many-particle
system under the influence of a self-consistent mean field potential and
a collision term. For the case of identical particles it is given by:
\begin{equation}
\label{buugl}
\frac{\partial f}{\partial t} + \frac{\vec{p}}{m}\, 
\frac{\partial f}{\partial \vec{r}}-
\vec{\nabla}U\, {\partial f \over \partial \vec{p}}=I_{coll}[f]
\quad,
\end{equation}
where $f(\vec{r},\vec{p},t)$ stands for the one-particle phasespace
density, $U[f]$ denotes the
self-consistent mean field potential and $I_{coll}[f]$ is the collision
term which - for a fermionic system - respects the Pauli principle.
For the description
of a system of non-identical particles one gets an equation for
each particle species that is coupled to all others by the
collision integral or the mean field potential.
Besides the nucleon we take all baryonic resonances up to
a mass of 2 GeV as well as the pion, the eta- and the rho-meson into account.
Schematically one can write down the set of coupled equations
in the following way:
\begin{eqnarray}
\label{gekbuu}
D\,f_N&=&F^N_{coll}(f_N,f_{\Delta(1232)},f_{N(1440)},\ldots,f_{\pi},f_{\eta},
\ldots)
\nonumber
\\
&=&I^N_{NN}+I^N_{N\Delta}+I^N_{NN(1440)}+\ldots+ I^N_{\Delta N \pi}+
I^N_{N(1440) N \pi}+ \ldots
\nonumber
\\
D\,f_{\Delta(1232)}&=&F^{\Delta(1232)}_{coll}(f_N,f_{\Delta(1232)},
f_{N(1440)},\ldots,f_{\pi})
\nonumber
\\
D\,f_{N(1440)}&=&F^{N(1440)}_{coll}(f_N,f_{\Delta(1232)},f_{N(1440)},
\ldots,f_{\pi},\,\ldots)
\nonumber
\\
\ldots&&
\nonumber
\\
D\,f_{\pi}&=&F^{\pi}_{coll}(f_N,f_{\Delta(1232)},f_{N(1440)},\ldots,f_{\pi},
\ldots)
\nonumber
\\
D\,f_{\eta}&=&F^{\eta}_{coll}(f_N,f_{N(1535)},f_{\eta})\quad,
\nonumber
\\
\ldots&&
\end{eqnarray}
where $D$ abbreviates the operator of the Vlasov-equation given by
the lhs of equation (\ref{buugl}).  
\par The initialization of $f_N$ for a nucleus at rest is done by using
a Woods-Saxon type density distribution:
\begin{equation}
\rho(r)=\rho_0 \left[ 1+{\rm e}^{\frac{r-r_0}{\alpha}} \right]^{-1}
\end{equation}
with the following parameters:
\[r_0=1.124\,{\rm fm}\,A^{1/3} \quad \alpha=\left(0.024\,A^{1/3}+0.29\right) 
{\rm fm} \quad.\]
In momentum space we make use of a local Thomas-Fermi approximation.
The local Fermi momentum $p_F(r)$ is thus given by:
\begin{equation}
\label{fermim}
p_F(r)=\sqrt[3]{\frac{3}{2} \pi^2 \, \rho(r)} \quad.
\end{equation}
\par The mean field $U_N$ for the nucleons consists of a Skyrme and a
Yukawa part and is described in \cite{cass90}. This potential is also
used for all isospin-$\frac{1}{2}$ resonances.
For the $\Delta$-resonance
we use a simple ansatz:
\begin{equation}
\label{delpot}
U_{\Delta}(\rho)=U_{\Delta,0}\,\frac{\rho}{\rho_0} \quad,
\end{equation}
with $U_{\Delta,0}=-30 \mbox{ MeV}$ \cite{ericsonweise}.
\par The collision term allows for the following reactions:
\begin{eqnarray*}
N\,N&\to&N\,N
\\
N\,N&\leftrightarrow& N\,R
\\
N\,N&\leftrightarrow& N\,N\,\pi \quad ({\rm S-wave})
\\
N\,R&\to&N\,R^{\prime}
\\
R&\leftrightarrow& N\,m
\\
R&\leftrightarrow& N\,\pi\,\pi
\\
&\leftrightarrow& \Delta(1232)\,\pi,\;N(1440)\,\pi,\;N\,\rho,\;
N\,\sigma
\\
\pi\,\pi&\leftrightarrow&\rho,\;\sigma \quad,
\end{eqnarray*}
where R stands for a baryonic resonance and m for a meson. The used
decay widths and cross sections are described in detail in
\cite{abspaper,teis}.
\section{Parameterization of the elementary $\gamma \, N$ 
\protect\linebreak cross sections}
\label{ele}
\subsection{One-pion production}
A coherent decomposition of the one-pion production cross sections into
resonance and background contributions is necessary since - especially in
the region of the $\Delta$-resonance - interference terms are quite
important. Therefore the resonance contributions are not fitted to total
cross sections but to partial-wave amplitudes $A_{l\pm}$ and $B_{l\pm}$
($j=l\pm \frac{1}{2}$, $l=$ angular momentum of the $\pi N$-system)
\cite{walker}. The amplitudes are taken from \cite{Arndt}.
\par For the resonance amplitudes we make - following \cite{walker} - a
Breit-Wigner ansatz under the assumption that the complex phase of the
amplitude is equal to
the phase of the resonance propagator for a spinless particle:
\begin{equation}
\label{walkerres}
\left\{
\begin{array}{c}
A_{R}(\sqrt{s})
\\
B_{R}(\sqrt{s})
\end{array}
\right \}
=
\left( \frac{k_0\,q_0}{k\,q} \right)^{1/2} \,
\left( \frac{\Gamma_0}{\Gamma_{\pi}(M_R)} \right)^{1/2} \,
\frac{\sqrt{s}\,\Gamma_{\pi}^{1/2}\,
\Gamma_{\gamma}^{1/2}}
{M_R^2-s-{\rm i}\,\sqrt{s}\,\Gamma_{tot}}
\left\{
\begin{array}{c}
A_{l\pm}(M_R) 
\\
B_{l\pm}(M_R) 
\end{array}
\right \}
,
\end{equation}
where
\begin{equation}
\label{ggamma}
\Gamma_{\gamma}=\Gamma_0 \left(\frac{k}{k_0} \right)^{j_1}
\left( \frac{k_0^2+X^2}{k^2+X^2} \right)^{j_2}
\end{equation}
and
\begin{eqnarray}
A_{l\pm}(M_R)&=&\mp \alpha \, C_{N\pi} \,A_{1/2} \\
B_{l\pm}(M_R)&=&\pm 4 \alpha \left[ (2J-1)(2J+3) \right]^{-1/2}\,
C_{N\pi}\,A_{3/2}
\end{eqnarray}
\begin{equation}
\alpha \equiv \left[ \frac{1}{\pi} \, \frac{k_0}{q_0} \, \frac{1}{2J+1} \,
\frac{M_N}{M_R} \, \frac{\Gamma_{\pi}(M_R)}{\Gamma_0^2} \right]^{1/2} \quad.
\end{equation}
Here $k$ and $q$ are the photon and pion 3-momentum in the cms for a
given cms energy $\sqrt{s}$. $k_0$ and $q_0$ are taken at the pole of
the resonance. $\Gamma_{\pi}$ denotes the one-pion decay width and $\Gamma_0$
the total vacuum width at the pole of the resonance.
$C_{N\pi}$ is the Clebsch-Gordan coefficient for the coupling of the isospins
of the pion and the nucleon to the isospin of the resonance.
The helicity amplitudes 
$A_{1/2}$ and $A_{3/2}$ are taken from \cite{PDB94}. For the parameter $X$
we use
\[X=0.3 \, {\rm GeV} \]
for all resonances. The parameters $j_1$ and $j_2$ are taken from
\cite{walker},
except for the $N(1535)$.
In table \ref{helitable} all parameters are listed for the resonances that
are important for photoabsorption.
\begin{table}[t]    
\centerline{
\begin{tabular}{|c|c|c|r|r|r|r|} 
\hline
& & &\multicolumn{2}{|c|}{ $A_{1/2}$ [${\rm GeV^{-1/2}}$]}&
\multicolumn{2}{|c|}{$A_{3/2}$ [${\rm GeV^{-1/2}}$]}\\
\cline{4-7}
\raisebox{1.5ex}[-1.5ex]{resonance} & 
\raisebox{1.5ex}[-1.5ex]{$j_1$} &
\raisebox{1.5ex}[-1.5ex]{$j_2$} &p&n&p&n\\
\hline
$P_{33}$(1232) & 2 & 1& -0.141 &  -0.141 & -0.260 & -0.260 \\
\hline
$D_{13}$(1520)    & 2 & 1& -0.022 &  -0.062  &  0.163   &  -0.137   \\
\hline
$S_{11}$(1535)    & 1 & 0 & 0.125   &  -0.100     &  0&  0 \\
\hline
$F_{15}$(1680)    & 4 & 2 & -0.014      & 0.027     & 0.135      & -0.035\\
\hline
\end{tabular}}
\caption{Resonance parameters for photoabsorption from \cite{PDB94,walker,
Kru95}.}
\label{helitable}
\end{table}
The resonance cross section corresponding to equation (\ref{walkerres}) is:
\begin{equation}
\label{walkerres2}
\sigma_{\gamma N \to R \to N \pi }=
\left(\frac{k_0}{k}\right)^2\, \frac{s\,\Gamma_{\gamma}\,\Gamma_{R \to N \pi }}
{\left( s-M_R^2 \right)^2 +s\,\Gamma_{tot}^2} \,
\frac{2M_N}{M_R\,\Gamma_0} \,
\left( \left|A_{1/2}\right|^2+\left|A_{3/2} \right|^2 \right) \, .
\end{equation}
The parameters of the $N(1535)$ were fitted to experimental eta
photoproduction data \cite{Kru95} that will be described in section \ref{eleta}.
The helicity amplitudes extracted from pion photoproduction \cite{PDB94} are
about a factor of 2 smaller than those in \cite{Kru95}. 
However, for the calculation of pion
photoproduction this is not important since the
contribution of the $N(1535)$ is anyway small.
\par
\begin{figure}[t]
\centerline{
\rotate[r]{\psfig{figure=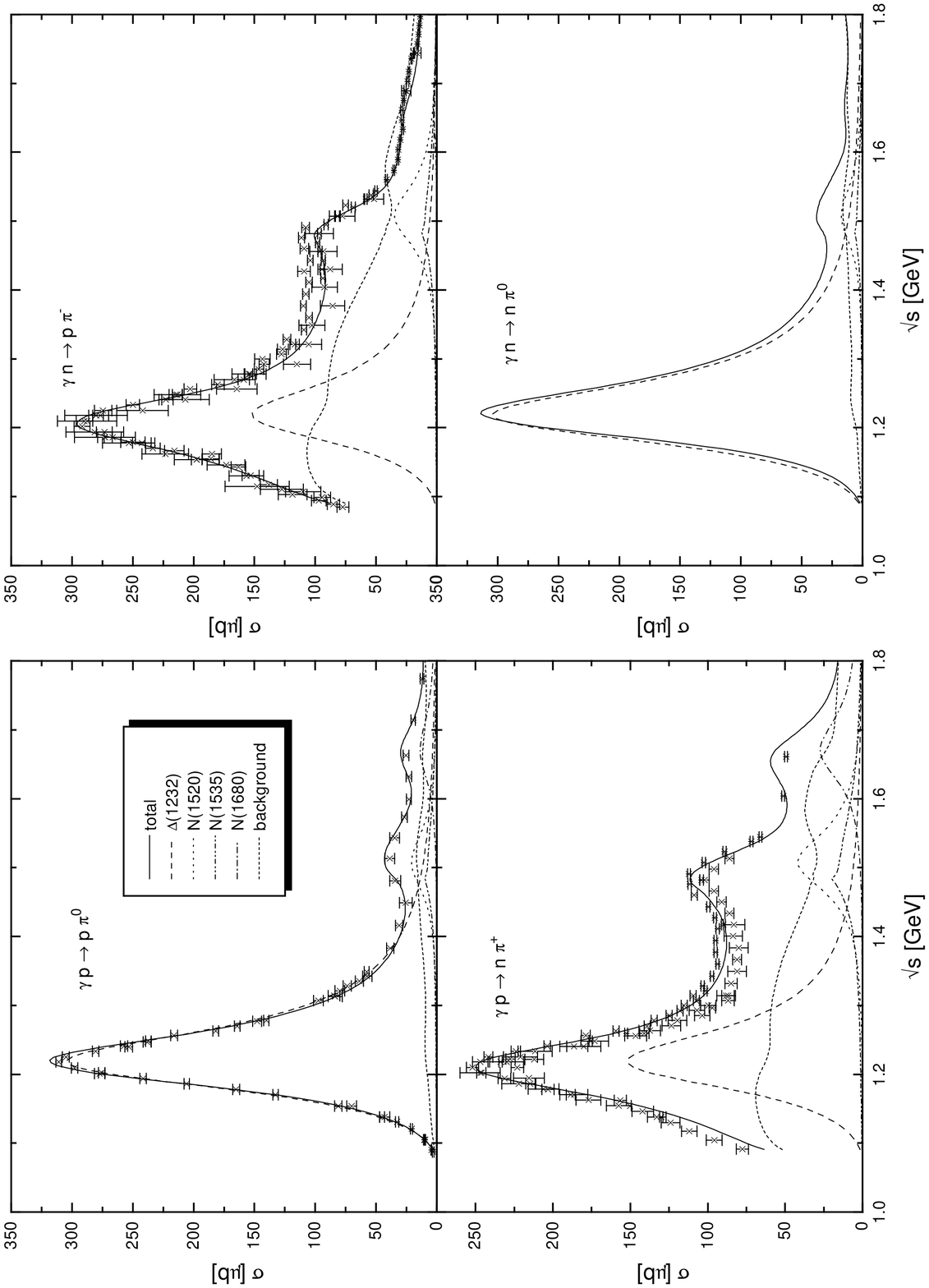,height=15cm}}}
\caption{Resonance and background contributions for one-pion photoproduction
cross sections
on the nucleon. The experimental data are from \protect\cite{eledaten}.}
\label{ele1}
\end{figure}

 In figure \ref{ele1} we show our decomposition of the total pion
photoproduction cross sections on the nucleon for all isospin channels.
The total cross section is obtained by summing coherently over the
individual resonance amplitudes as well as a background amplitude.
The curve labeled 'background' represents the cross section that is obtained after
subtracting all resonance contributions 
(cf. eq. (\ref{walkerres}))
from the partial-wave 
amplitudes \cite{Arndt}.
The smooth background indicates that our procedure is reasonable.
\par An incoherent addition of all contributions, as
performed by Kondratyuk et al. \cite{kondr94},
differs significantly from the coherent one, especially in the 
$\Delta$-region for the $\pi^+$-channel.
A coherent summation of the resonance and background contributions is thus
mandatory if one wants to investigate possible modifications of the
resonance contributions in nuclei.
\par Since there are no experimental data for the
$\pi^0$-cross section on the neutron this channel suffers from the
uncertainties in the partial-wave analysis from \cite{Arndt}.
However, since the non-resonant background in this channel is 
- at least in the region of the $\Delta$-resonance - expected
to be small \cite{ericsonweise} the cross section consists
mainly of resonance contributions that are fixed by measurements of the
other channels and isospin symmetry. The uncertainties of the helicity
amplitudes of the $\Delta$-resonance are only of the order of a few
percent \cite{PDB94}. The used cross section is also in line with an
extraction from the deuteron cross section \cite{Kruschepi0}. Therefore
we expect that our parameterizations of the one-pion photoproduction cross
sections are reasonable.
\subsection{Etaproduction}
\label{eleta}
\begin{figure}[t]
\centerline{
\rotate[r]{\psfig{figure=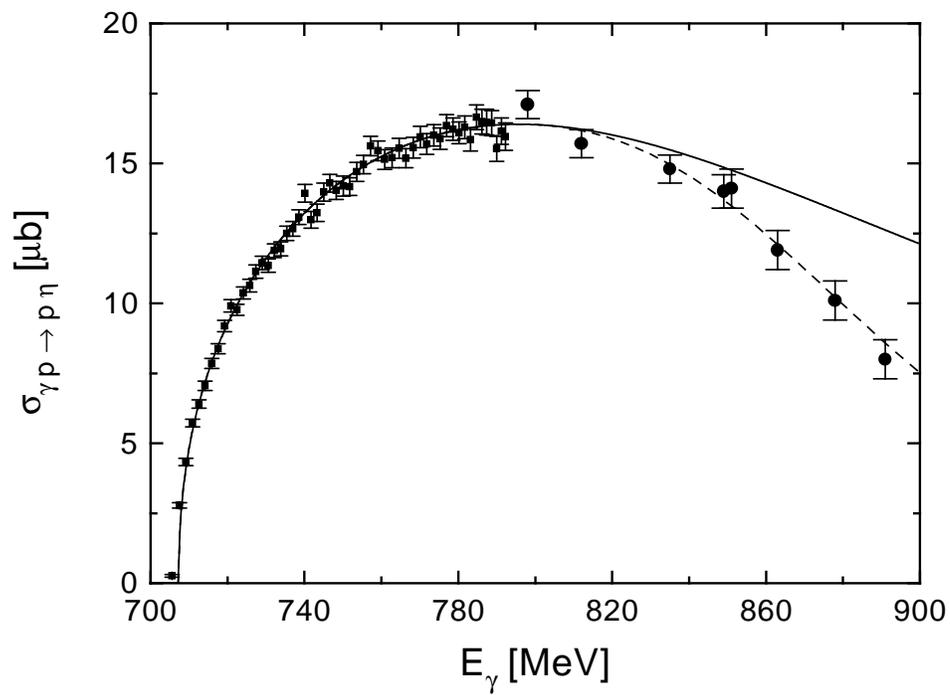,height=15cm}}}
\caption{Cross section for eta photoproduction on the proton. The
experimental data are taken from \protect\cite{Kru95} (squares) and 
\protect\cite{elsa} (circles). Explanation see text.}
\label{eletag}
\end{figure}

The cross section for eta photoproduction is parameterized under the
assumption that the only production mechanism is that through the excitation 
of an intermediate $N(1535)$-resonance \cite{Kru95}.
Then we get analogously to equation (\ref{walkerres2}) the cross section
within a Breit-Wigner approximation:
\begin{equation}
\sigma_{\gamma N \to N \eta}=\left(\frac{k_0}{k} \right)^2 
\frac{s\,\Gamma_{\gamma} \,\Gamma_{N(1535) \to N \eta}}
{\left(s-M_{1535}^2\right)^2+s\,\Gamma_{tot}^2}
\, \frac{2M_N}{M_{1535}\,\Gamma_0}\,\left|A_{1/2}\right|^2 \quad,
\end{equation}
where $\Gamma_{\gamma}$ has been defined in equation (\ref{ggamma}). The
total width $\Gamma_{tot}$ is given by:
\begin{equation}
\Gamma_{tot}=\Gamma_{\pi}+\Gamma_{\eta}+\Gamma_{\pi\pi} \quad,
\end{equation}
with
\begin{eqnarray}
\Gamma_{\pi}&=&\frac{q_{\pi}}{q_{R,\pi}}\,b_{\pi}\,\Gamma_0\,
\frac{q_{R,\pi}^2+c_{\pi}^2}{q_{\pi}^2+c_{\pi}^2}\\
\Gamma_{\eta}&=&\frac{q_{\eta}}{q_{R,\eta}}\,b_{\eta}\,\Gamma_0\,
\frac{q_{R,\eta}^2+c_{\eta}^2}{q_{\eta}^2+c_{\eta}^2}\\
\Gamma_{\pi\pi}&=&b_{\pi\pi} \, \Gamma_0 \,\Theta(\sqrt{s}-M_N-2m_{\pi}) \quad.
\end{eqnarray}
The position of the resonance pole $M_{1535}$ and the total width $\Gamma_0$
with the branching ratios
$b_{\pi}$, $b_{\eta}$ and $b_{\pi\pi}$ as well as the helicity amplitude
$A_{1/2}$ are adopted from \cite{Kru95}. The fit to the experimental
data was improved by multiplying form factors with parameters $c_{\pi}$ and
$c_{\eta}$ to the partial widths. With
\[c_{\pi}^2=c_{\eta}^2=0.25\, {\rm GeV^2} \]
one obtains the solid curve shown in figure \ref{eletag}. For photon 
energies larger than 800 MeV only electroproduction data with a
low virtual photon momentum transfer of $q^2=-0.056\,{\rm GeV^2}$
are available \cite{elsa}. Following \cite{kaiser} where it is argued that these data can 
be regarded as photoproduction data within the errorbars, we will also use a
different
fit including these data points (dashed curve in figure \ref{eletag}) for
comparison. This will only affect the electromagnetic properties of
the $N(1535)$-resonance, but not the hadronic ones.
\par For the neutron cross section we use the
proton cross section scaled by a factor 0.64 according to \cite{Krudeut}.
\subsection{Two-pion production}
\begin{figure}[t]
\centerline{
\psfig{figure=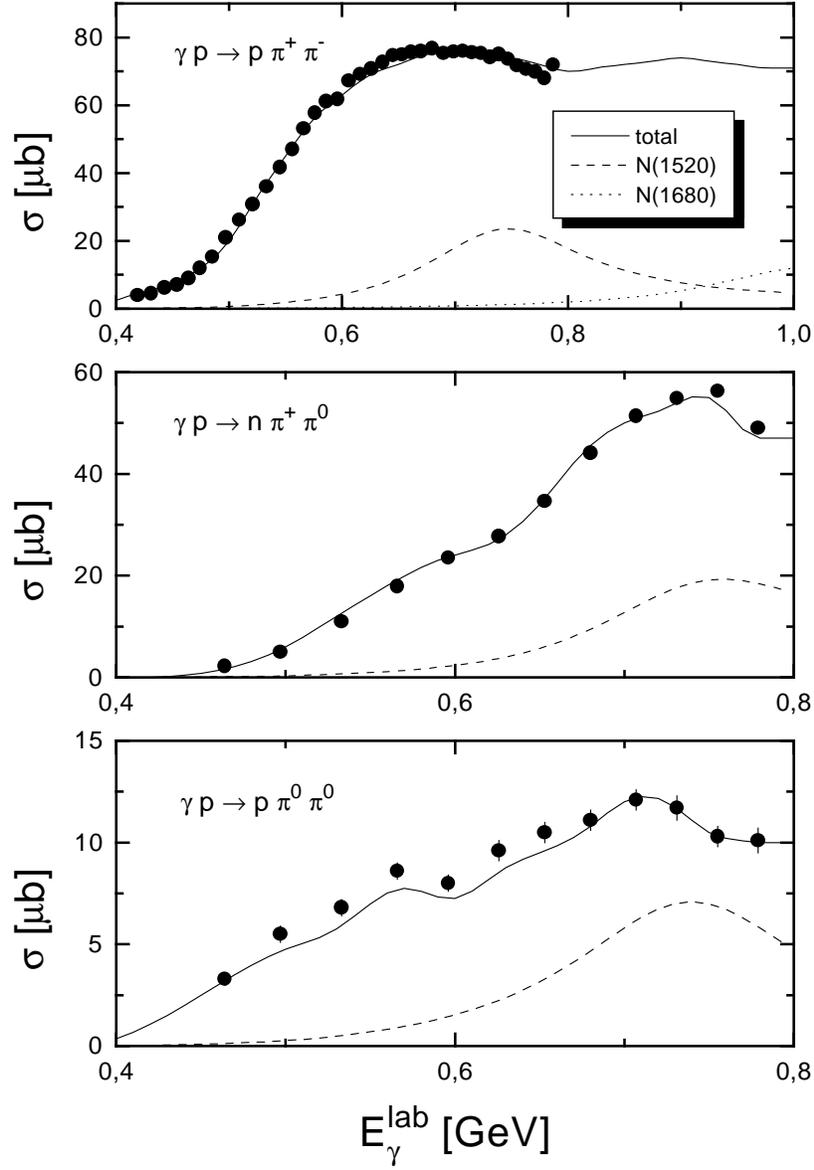,width=12cm}}
\caption{Cross sections for two-pion photoproduction on the proton. The
experimental data are from \protect\cite{brag}.} 
\label{2pia}
\end{figure}

In figure \ref{2pia} we compare the contributions of the $N(1520)$ and
the $N(1680)$ with the measured two-pion photoproduction cross sections
on the proton for the three isospin channels. The resonance contributions
are given by the Breit-Wigner formula in equation (\ref{walkerres2}) with
$\Gamma_{R \to N \pi}$ replaced by $\Gamma_{R \to N\pi\pi}$. It is obvious
that these contributions explain neither the size nor the structure of the
cross section for any channel.
\par There are calculations of the
two-pion photoproduction cross section by L. Y. Murphy et al. \cite{Murphy95}
and J. A. Gomez Tejedor et al. \cite{Oset94} within the frame\-work of a
phenomenological pion nucleon field theory.
In both calculations the number of diagrams giving important contributions
is rather large with, unfortunately, only little agreement in the
assignment of the most important ones. Therefore one can expect insight
into the underlying processes only by the experimental measurement of
massdifferential
cross sections. Up to then we treat the difference between the
experimental cross section and the Breit-Wigner type resonance contributions
as background, where the momenta of the outgoing particles are distributed
according to three-body phase space. Therefore the only medium modification is
the possible Pauli blocking of the outgoing nucleon. This approach is
somewhat unsatisfactory since the 'background' is probably dominated more by
intermediate excitations of baryonic and mesonic resonances than by
contributions coming from diagrams involving pions and nucleons only.
For photon energies that exceed the measured energy range we assume the
cross sections to be constant.
\par Because there are no recent data available for $2 \pi$ production on 
the neutron we obtain these from
the difference between the total photoabsorption cross section and the
one-pion photoproduction cross section. We use an ansatz for the
background contribution:
\begin{equation}
\sigma_{\gamma n \to N\pi\pi}^{bg}=\frac{a_1\,x+a_2\,x^2}
{1+a_3\,x+a_4\,x^2}\,{\rm \mu b}
\end{equation}
\[x=\frac{\sqrt{s}-M_N-2m_{\pi}}{{\rm GeV}}\]
and obtain for the fit parameters:
\[a_1=249.7 \quad a_2=314.1 \quad a_3=-0.3537 \quad a_4=4.604 \quad.\]
The decomposition into the different isospin channels is done by using
the maximum values of the two-pion cross sections on the proton. This
gives independent of photon energy:
\[
\sigma_{\gamma n \to n \pi^+ \pi ^-}:\sigma_{\gamma n \to p \pi^- \pi^0}:
\sigma_{\gamma n \to n \pi^0 \pi^0}=75:55:12 \quad.
\] 
\subsection{Implementation into the BUU model}
In the BUU model nucleon resonances are explicitly propagated in order to
account for multi-step processes.
Therefore we replace the coherent sum for the one-pion
photoproduction cross section by a properly scaled incoherent cross section. This gives an
effective cross section for the photoproduction of a resonance R:
\begin{equation}
\sigma^{eff}_{\gamma N \to R}=\int  
\frac{\frac{d\sigma_{\gamma N \to R \to N \pi}}{d\Omega}}     
{\frac{d\sigma^{bg}_{\gamma N \to N \pi}}{d\Omega} + 
\sum_{R^{\prime}}
\frac{d\sigma_{\gamma N \to R^{\prime} \to N \pi} }{d\Omega} }\,
\frac{d\sigma_{\gamma N \to N \pi}}{d\Omega}
\,d\Omega
+ \sigma_{\gamma N \to R \to X (X \neq N \pi)} \quad,
\end{equation}
where $\sigma_{\gamma N \to R \to N \pi}$ and
$\sigma^{bg}_{\gamma N \to N \pi}$ denote the cross sections without
interferences. The sum $\sum_{R^{\prime}}$ runs over all contributing
resonances. Analogously the cross section for the direct production of
a $N\pi$-pair is given by:
\begin{equation}
\label{npibackground}
\frac{d\sigma^{bg,eff}_{\gamma N \to N \pi}}{d\Omega}=
\frac{\frac{d\sigma^{bg}_{\gamma N \to N \pi}}{d\Omega}}     
{\frac{d\sigma^{bg}_{\gamma N \to N \pi}}{d\Omega} + 
\sum_R
\frac{d\sigma_{\gamma N \to R \to N \pi} }{d\Omega} }\,
\frac{d\sigma_{\gamma N \to N \pi}}{d\Omega} \quad.
\end{equation}
This procedure guarantees that the sum of these cross sections is equal
to the cross section obtained by coherent addition of all contributions.
\par From figure \ref{ele1} one sees that for the channels
with charged pions the size of the interference terms of resonance and
background contributions amounts to at most about 20\% of the total
one-pion cross sections. Since for all resonances the width for spontaneous
decay is much larger than the collision width (see \cite{abspaper}) the 
uncertainties induced by our decomposition
of the cross section into a sum of incoherent contributions are
negligible.
\par The medium modifications for the elementary cross sections are described
in \cite{abspaper}. The vacuum width appearing in the resonance propagator in
equation (\ref{walkerres}) is replaced by the in-medium width as calculated
in \cite{abspaper}. The collision width gives a contribution to the absorption cross
section of the form of equation (\ref{walkerres2}) with $\Gamma_{R,coll}$
instead of $\Gamma_{R \to N \pi}$. For the $\Delta$-resonance the difference between nucleon and
$\Delta$-potential (equation (\ref{delpot})) causes a real part
of the self energy $\Pi$ to be used in the resonance propagator:
\begin{equation}
{\rm Re}\,\Pi=2\,E_{\Delta}\,\left( U_N-U_{\Delta} \right) \quad.
\end{equation}
We also take into account that nucleon final states can be Pauli blocked. 
The decay of the resonances is, for simplicity, always assumed to 
be isotropic in the resonance rest frame.
\section{Results}
\subsection{Pion photoproduction}
\label{pionprod}
\begin{figure}[t]
\centerline{
\psfig{figure=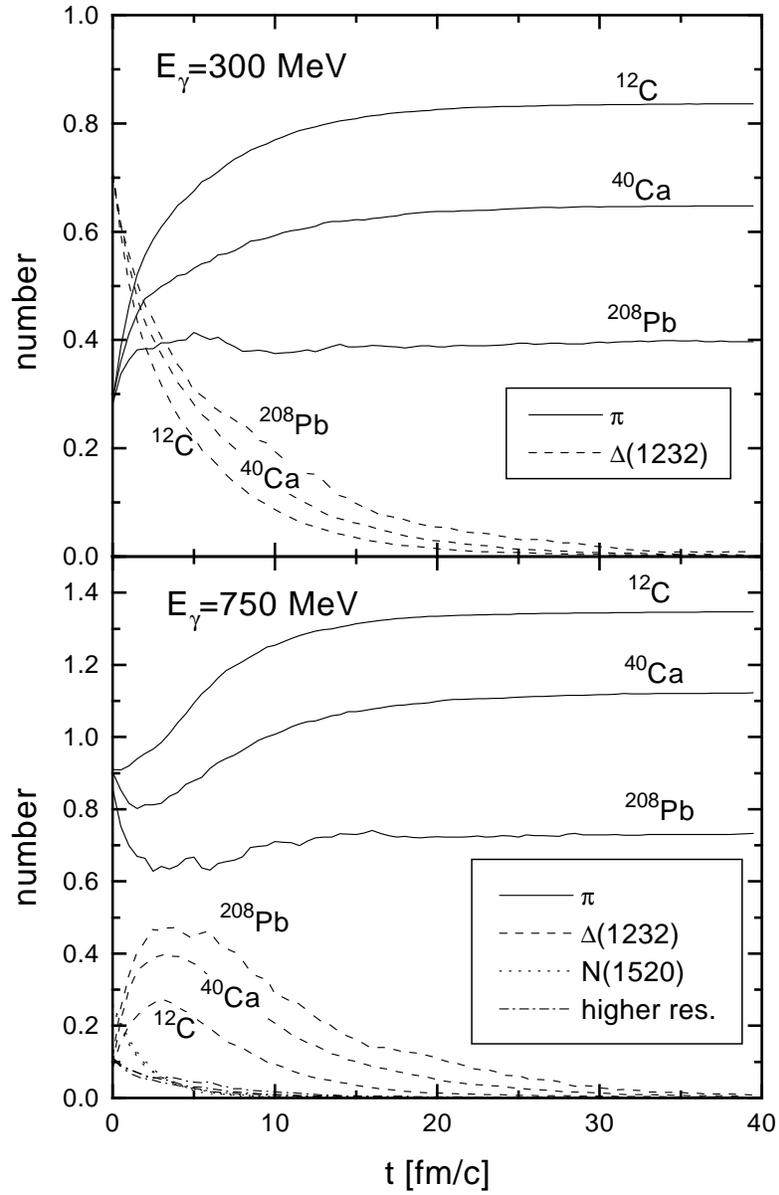,width=11.5cm}}
\caption{Time evolution of the photon-nucleus reaction. The curves represent
average numbers of particles (integrated over isospin) per absorbed photon 
in a nucleus.} 
\label{time}
\end{figure}

In figure \ref{time} we show the time evolution of the photon nucleus
reaction for different nuclei and photon energies of 300 and 750 MeV. The
duration of the reaction is about 20 fm/c and more or less independent of
the photon energy. For both photon energies the $\Delta$-resonance plays
an important role,
but its time-development is quite different. Whereas at 
$E_{\gamma}=300$ MeV the main $\gamma$ absorption goes through the
$\Delta$ which subsequently decays, at 
750 MeV only a small number of $\Delta$'s is directly
excited by the photon. Most of the $\Delta$'s are populated by collisions
of nucleons with pions from the absorption process
$\gamma \, N \to N\, \pi \, \pi$. From the decay time of the
$\Delta$-contribution, being much larger than the lifetime of a single
$\Delta$, one can see that the reaction is indeed a multi-step process
dominated by continous excitation and decay of $\Delta$-resonances. The
excitation propagates alternately as $\Delta$ or pion through the nuclear
medium.
\par
\begin{figure}[t]
\centerline{
\psfig{figure=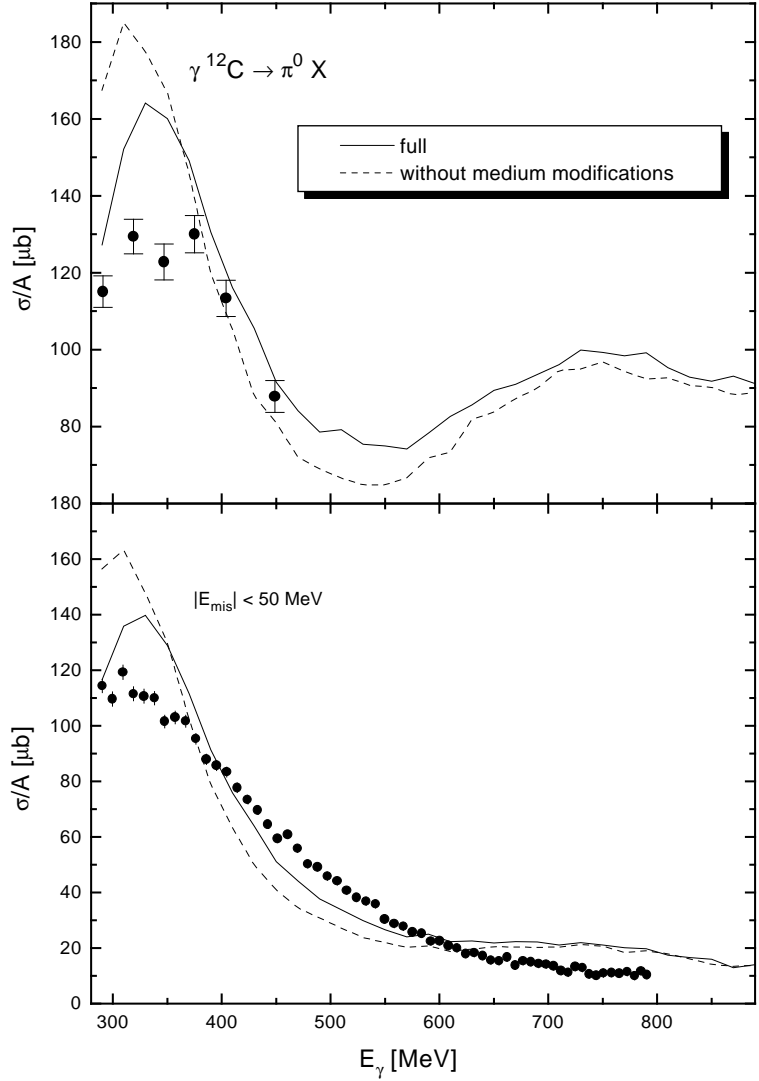,width=10.5cm}}
\caption{$\pi^0$ photoproduction cross sections on $^{12}$C.
The upper figure shows the total cross section
(experimental data are from \protect\cite{arends}). The lower figure depicts
the cross section with a cut on the energy of the $\pi^0$ (see text, 
experimental data from \protect\cite{Kruschepi2}). The dashed curves
result when using the vacuum widths for the resonances and neglecting
the difference between nucleon- and $\Delta$-potential.}
\label{ctpi}
\end{figure}

 The total $\pi^0$ cross section on $^{12}$C is shown in figure \ref{ctpi}.
The curve labeled 'without medium modifications' results when applying only
Fermi motion and Pauli blocking and neglecting the difference between  
nucleon- and $\Delta$-potential and
the in-medium widths of the resonances. The difference to the full
calculation is basically caused by the $\Delta$-potential. In the
region of the $\Delta$-resonance the elementary absorption cross
section is reduced and shifted to higher energies \cite{abspaper}. For
larger photon energies the $\Delta$-potential leads to a slight
enhancement of the pion production due to its effect on the pion-nucleon
interaction.
Compared to the experimental data from \cite{arends} our calculation
overestimates the cross section in the region of the $\Delta$-resonance
by about 30\%. In recent
experiments in Mainz \cite{Kruschepi2} only pions within a 'missing energy'
window were detected. The cut on the 'missing energy' $E_{mis}$ led to
a preferred detection of quasi-free produced pions which had no final state
interaction. In the pole region of the $\Delta$-resonance we again
overestimate the cross section by about the same factors as the total cross
section and at the high energy tail we underestimate it but here the 
cross section is too small to draw any firm conclusions. 
\par
\begin{figure}[t]
\centerline{
\psfig{figure=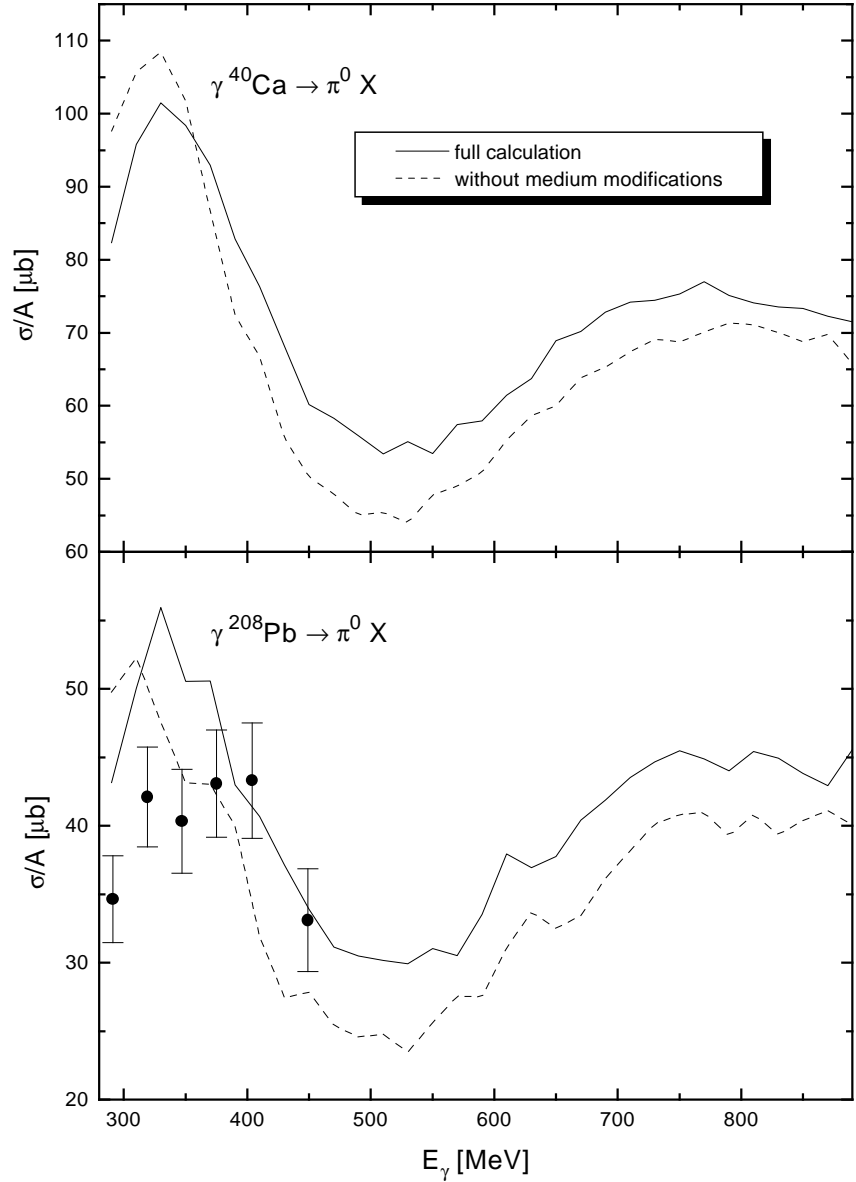,width=11.5cm}}
\caption{Total $\pi^0$ photoproduction cross sections on
$^{40}$Ca and $^{208}$Pb. The experimental data are taken from \cite{arends}.
The fluctuations in the calculated curves are caused by low statistics.}
\label{catpi}
\end{figure}

 In figure \ref{catpi} we show our results on $^{40}$Ca and $^{208}$Pb. 
Qualitatively the curves look similar to the ones for $^{12}$C.
The cross section on $^{208}$Pb shows somewhat better
agreement with the experimental data than in the case of $^{12}$C,
but we again fail to explain the broad structure in the $\Delta$-region. 
\par The photoproduction of charged pions in nuclei was measured only up to
photon energies of 400 MeV \cite{arends82}. Similar to the $\pi^0$ production
we overestimate these cross sections by about 25\%.
\par The discrepancy in the $\Delta$-region might be due to a further reduction
of the cross section for $\gamma \,N \to N \,\pi$ in the nuclear medium.
A better description of the total photoabsorption cross section
then certainly requires 
the inclusion of two- and three-body absorption mechanisms
for the photon. This may also lead to a better description of the observed
structure of the cross sections.
However, one has to note that only a part of all produced pions actually
survives the final state absorption on their way out of the nucleus. Thus,
one needs a very reliable description of pion-absorption in order to
become sensitive to modifications of the elementary photon-nucleon 
interaction.
\par
\begin{figure}[t]
\centerline{
\rotate[r]{\psfig{figure=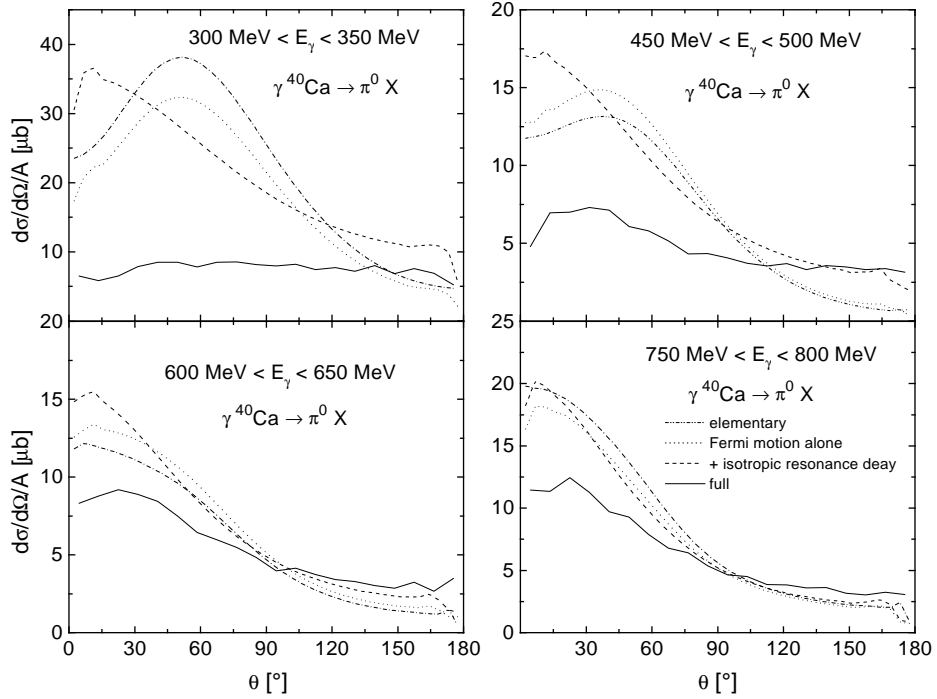,height=15cm}}}
\caption{Angular differential $\pi^0$ photoproduction cross sections on
$^{40}$Ca for different photon energies. The dash-dotted curves show the
cross sections on a free nucleon. Applying Fermi motion alone leads to the
dotted curves. An isotropic resonance decay, which is assumed in the
BUU calculation, gives the dashed curves (without final state interactions
of the outgoing pions). The solid lines represent the full BUU calculation.}
\label{capwi}
\end{figure}

 The shape of the energy and angular differential cross sections is
nearly independent of the mass of the nucleus. Therefore we limit ourselves
here to the discussion of the results for $^{40}$Ca. In figure \ref{capwi}
angular differential cross sections for the production of $\pi^0$'s are
shown for different photon energies. 
From the difference between the curves labeled 'Fermi motion alone' and
'+ isotropic resonance' one sees that
for photon energies less than 500 MeV, where the p-state interaction 
dominates, the
isotropic resonance decay, assumed in the BUU model, leads to an incorrect implementation of
the elementary angular differential cross sections. 
However, the final state interaction of the pions with the nuclear medium leads in general to
a higher isotropy of the cross section so 
that the final result is not affected by this deficiency.
For small photon energies the cross
section at forward angles is additionally reduced since a pion emitted at
forward angles corresponds to a low momentum nucleon which has a high
probability to be Pauli blocked. For higher photon energies this becomes
more and more unimportant and the cross section is shifted to forward
angles because of kinematical reasons. 
\par
\begin{figure}[t]
\centerline{
\rotate[r]{\psfig{figure=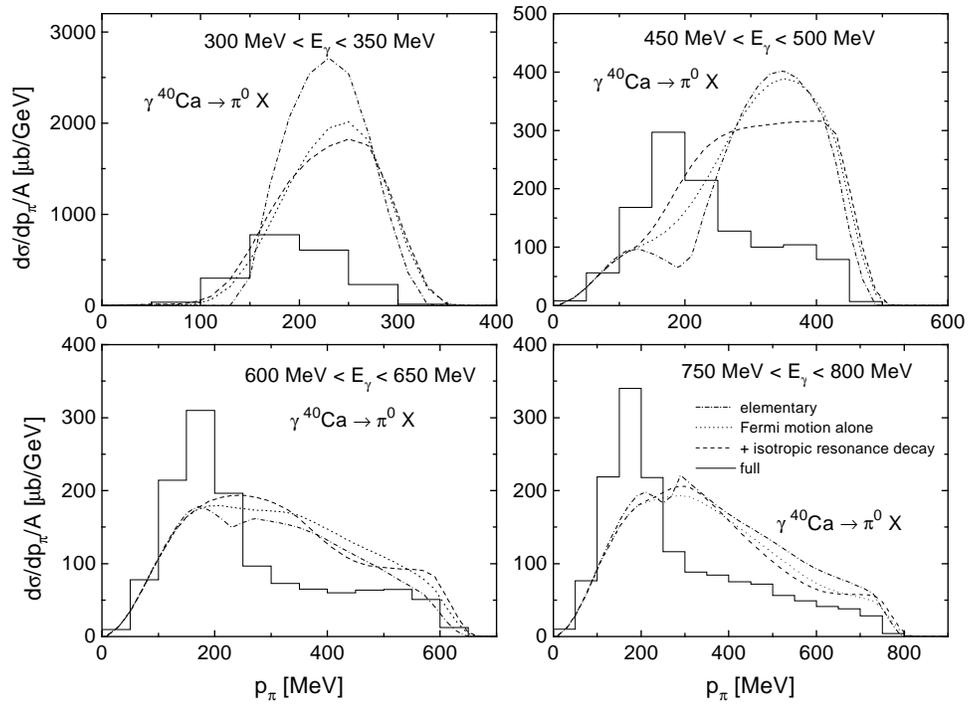,height=15cm}}}
\caption{Momentum differential $\pi^0$ photoproduction cross sections on
$^{40}$Ca for different photon energies. Explanation see figure \ref{capwi}.}
\label{capim}
\end{figure}

 Momentum differential cross sections for the same energy ranges of the
photon are given in figure \ref{capim}. Here the influence of the isotropic
resonance decay is less important. The structure in the elementary cross
section between photon energies of 450 and 650 MeV coming from the one-pion
and two-pion contribution is already washed out by Fermi motion. Independent
of photon energy one observes an enhancement of the cross section at pion
momenta of 200 MeV and a reduction at 400 MeV. This is due to the strong
final state
coupling of the pions to nucleons as discussed earlier. For smaller
pion momenta the pion nucleon cross section, dominated here by the
$\Delta$-resonance, gets smaller. Therefore pions 
with momenta around 200 MeV
have a higher
probability to escape from the nucleus. 
\par In figures \ref{capwi} and \ref{capim} it is noticeable that the
$\pi^0$ cross section 
compared to the elementary cross section is 
in the $\Delta$-region ($E_{\gamma}=300-350 \; {\rm MeV}$) much smaller 
than for higher energies. This is due to
charge exchange reactions by primary produced charged pions, whose
relative contribution to the elementary cross section increases with
increasing photon energy.
\par An experimental measurement of pion photoproduction in nuclei in the
energy range from 400 MeV to 1 GeV would be very helpful because with
increasing photon energy multi-body absorption mechanisms are expected to
become smaller. Then a comparison between theory and experiment would allow
cleaner conclusions about medium modifications of the elementary
photon-nucleon interaction.
\par
\begin{figure}[t]
\centerline{
\psfig{figure=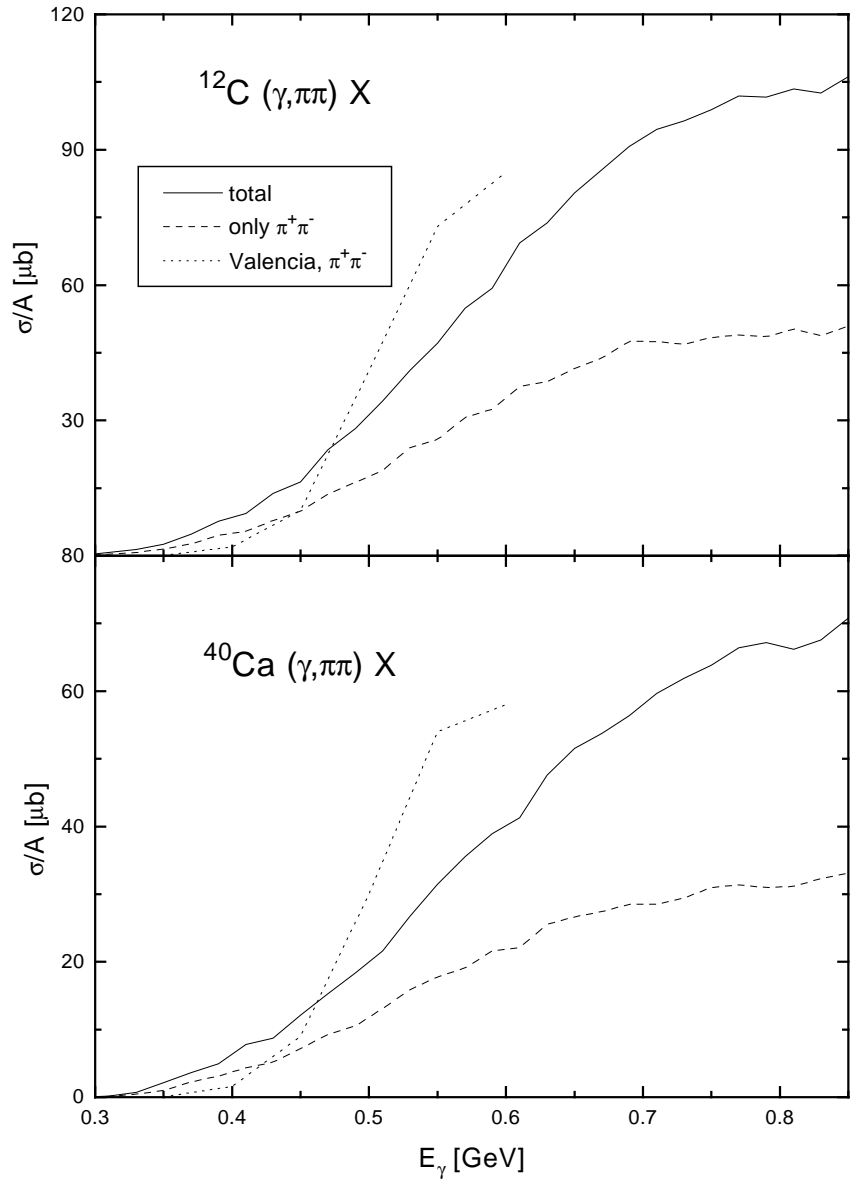,width=11.5cm}}
\caption{Total two-pion photoproduction cross section on $^{12}$C
and $^{40}$Ca. The
curve labeled 'Valencia' is taken from a calculation of J. A. Gomez
Tejedor et al. \protect\cite{oset2pikern}.}
\label{2pic12}
\end{figure}

 Figure \ref{2pic12} shows the inclusive two-pion
photoproduction cross sections on $^{12}$C and $^{40}$Ca. Since there are yet
no experimental data available we compare our results with the calculations of
J. A. Gomez Tejedor et al. \cite{oset2pikern} done
in the framework of a $\Delta$-hole model. The difference to our
calculation is due to the fact that in our calculation there is only a
small medium modification of the elementary $\gamma \, N \to N \, \pi \, \pi$
cross section.
In \cite{oset2pikern} this cross section is calculated microscopically using
medium modifications for the propagators and potentials for the phasespace
integrations that cause a strong enhancement of the elementary
cross section in the nuclear medium.
The differences between the two $\pi^+ \pi^-$ curves are so large that they
should be experimentally observable. Such a measurement would also
be very helpful with respect to the disappearance of the $D_{13}$ in the
total photoabsorption cross section \cite{abspaper}.
\subsection{Etaproduction}
\label{etaprod}
In section \ref{eleta} we have parameterized the etaproduction on the free
nucleon under the assumption that the only production mechanism is
via an intermediate $N(1535)$-resonance. In nuclei we also have to take into
account eta production by final state interactions of pions that
were primary produced.
\par
\begin{figure}[t]
\centerline{
\rotate[r]{\psfig{figure=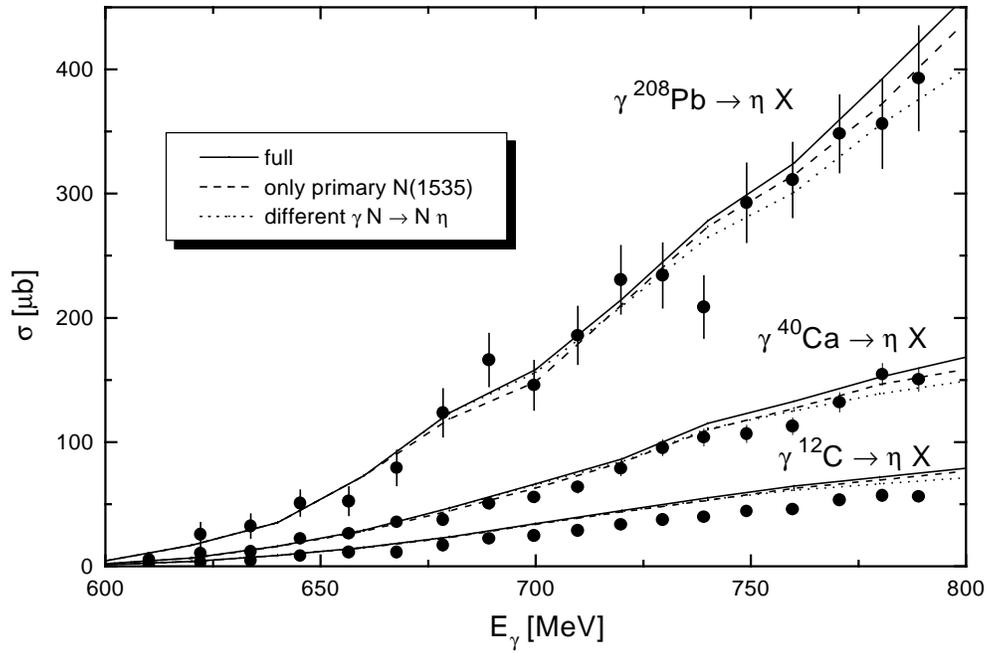,height=15cm}}}
\caption{Total eta photoproduction cross section on $^{12}$C,
$^{40}$Ca and $^{208}$Pb. All
experimental eta photoproduction data are taken from \protect\cite{robig-landau}.}
\label{calleta}
\end{figure}

 In figure \ref{calleta} we compare the calculated total etaproduction cross
section on $^{12}$C, $^{40}$Ca and $^{208}$Pb with experimental data from \cite{robig-landau}. The
contributions coming from secondary processes are almost negligible.
For large photon energies our calculation depends on the choice of the
elementary $\gamma \,N \to N \,\eta$ cross section for photon energies
larger than 800 MeV because of the Fermi motion of the nucleons. An extrapolation
of this cross section according to low momentum transfer electroproduction
data as discussed in chapter \ref{eleta} (dashed curve in figure
\ref{eletag}) reduces the total cross section at 800 MeV by about 15\% and
gives a better description of the experimental data.
On $^{12}$C there is only good agreement with the
experiment at low photon energies. For higher energies we overestimate
the cross section by about 20\%.
On $^{40}$Ca and $^{208}$Pb the agreement with the experiment is very good.
\par
\begin{figure}[t]
\centerline{
\rotate[r]{\psfig{figure=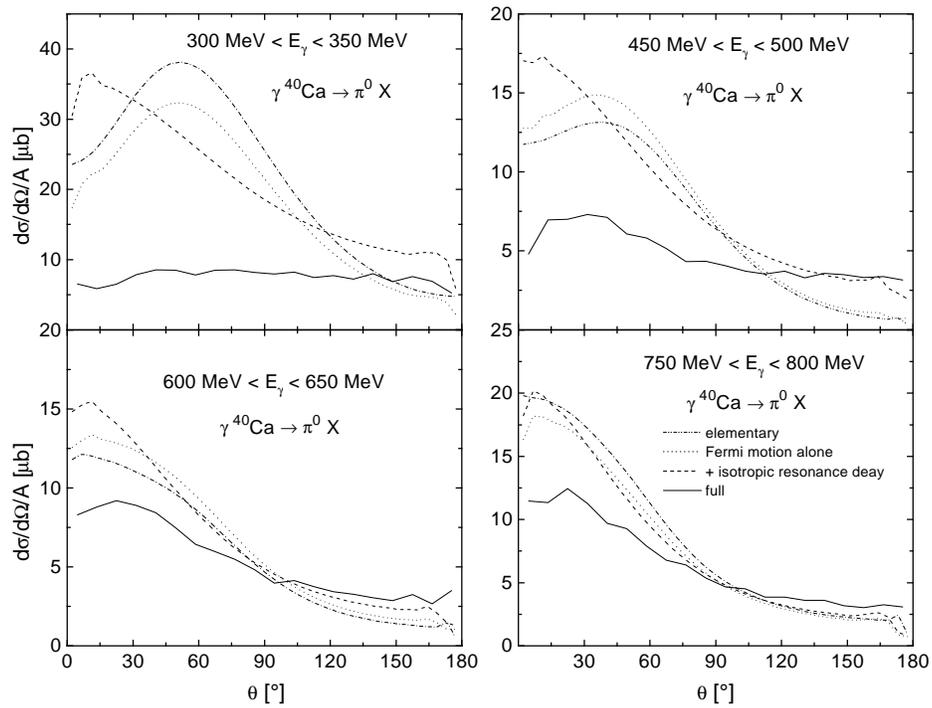,height=15cm}}}
\caption{Angular differential eta photoproduction cross sections on
$^{40}$Ca. The dashed curves result when applying Fermi motion alone
without any final state interaction of the outgoing etas. The dotted lines
represent the cross sections on a free nucleon at rest.}
\label{etang}
\end{figure}

 Angular differential cross sections on $^{40}$Ca are shown together 
with the
experimental data in figure \ref{etang} for different photon energies.
For comparison we also show the cross sections on a free nucleon and the
cross sections that result when applying Fermi motion alone and neglecting
any medium effects on the N(1535)-resonance.
The influence of the final state interaction on the shape of the angular
differential cross
sections is small. Compared to the data the calculated cross sections
are slightly shifted to smaller angles. The same holds for $^{12}$C and $^{208}$Pb.
\par
\begin{figure}[t]
\centerline{
\rotate[r]{\psfig{figure=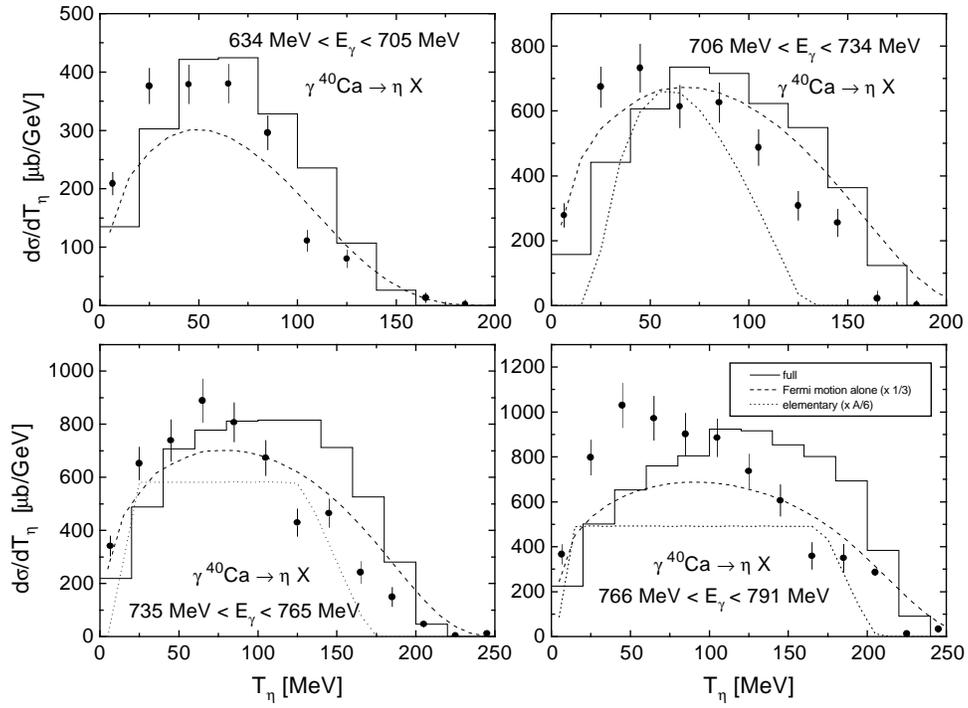,height=15cm}}}
\caption{Energy differential eta photoproduction cross sections on
$^{40}$Ca. Explanation see figure \ref{etang}.}
\label{etaim}
\end{figure}

 A corresponding discrepancy can be seen in energy differential cross
sections given in figure \ref{etaim}. Here the discrepancy is larger; 
the calculated cross sections are shifted
to larger eta energies since etas emitted at forward angles
have larger energies.
\par
\begin{figure}[t]
\centerline{
\psfig{figure=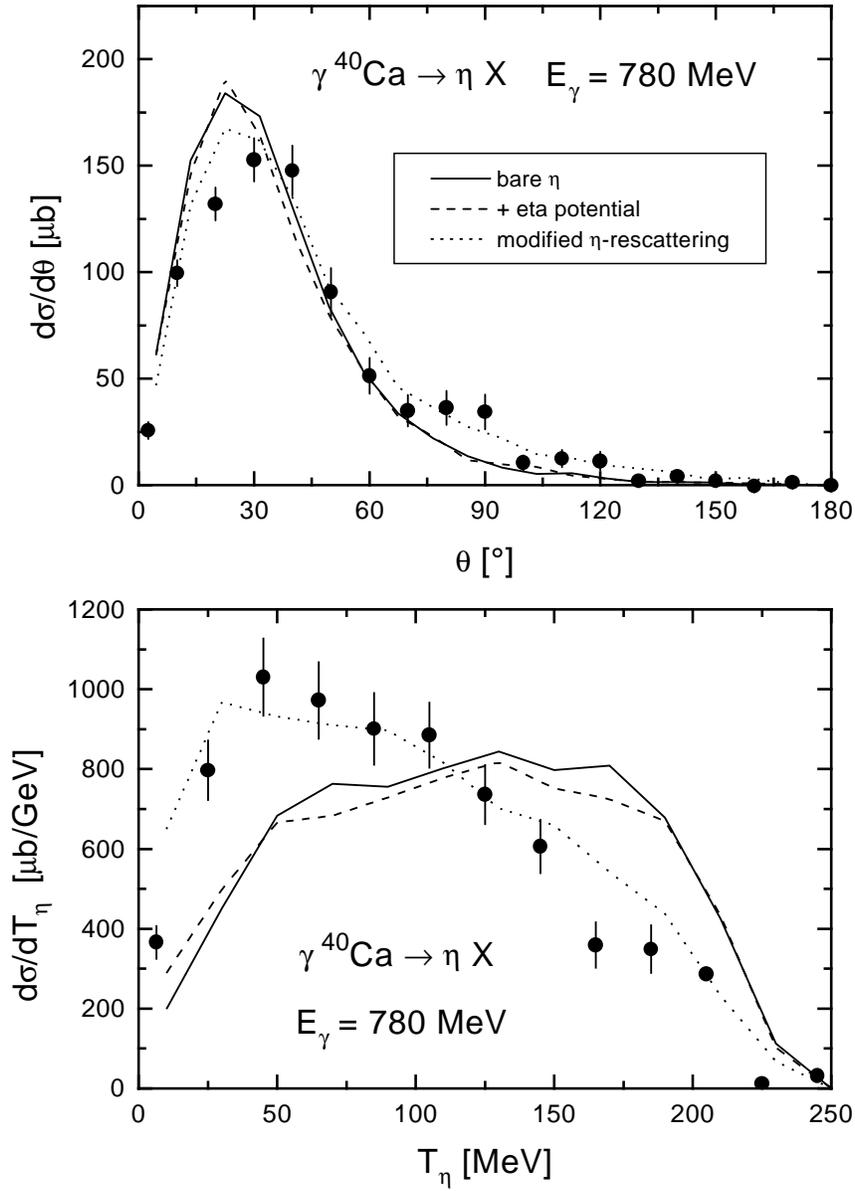,width=11.5cm}}
\caption{Modifications of the angular and energy differential eta photoproduction
cross sections on $^{40}$Ca. The solid lines represent the previous cross
section. Explanation see text.} 
\label{capwi2}
\end{figure}

 Possible reasons for the discrepancies in the shape of the observed 
differential cross
sections are an eta potential and a modification of the cross
section for eta-nucleon scattering. 
In order to check these possibilities we have first used
an eta potential as given by Lee et al. \cite{lee}.
This potential is attractive for eta energies below 100 MeV (-30 MeV for
$T_{\eta}=0$ at $\rho_0$) and slightly repulsive for larger energies. In our
calculation the potential enters the phase space integration for the eta
production as well as the equations of motions for the propagation of the
eta through the nucleus. In figure \ref{capwi2} the resulting differential
cross sections on $^{40}$Ca for a photon energy of 780 MeV are shown with
the previous obtained cross section. For simplicity we now neglect
contributions from primary produced pions or resonances other than N(1535)
since they are anyway small. From figure \ref{capwi2} one sees that the
effect of the potential is quite small.
\par Since only a small fraction of the etas produced
inside the nucleus survives the propagation through the nucleus (about
$1/3$ for $^{40}$Ca, $1/6$ for $^{208}$Pb) there is a strong dependence
of the eta production cross sections on
modifications of the eta absorption processes. 
Indeed we find that we are able to reproduce the experimental data by using a
constant inelastic $\eta \,N$-cross section, $\sigma_{in.}^{\eta}=30\,{\rm mb}$,
and a constant elastic cross section, $\sigma_{el.}^{\eta}=20\,{\rm mb}$. The 
resulting cross sections are labeled 'modified $\eta$-rescattering' in figure
\ref{capwi2}. One should note that energy and angular differential cross 
sections are then reproduced simultaneously. 
\section{Summary and outlook}
We have presented a calculation of the photoproduction of pions and
etas within the
framework of a semi-classical BUU transport model for photon energies
from 300 to 900 MeV. Starting from a realistic parameterization of the
free photon-nucleon cross sections we have applied the medium modifications
Fermi motion, Pauli blocking and collision broadening for the involved
nucleon resonances.
\par In the $\Delta$-region we are able to reproduce the size of the observed
pion production in nuclei reasonably well 
although our calculated cross sections show too much structure in the
resonance region.
This might be due to the importance of multi-body
absorption mechanisms. Our results for two-pion production differ by about
a factor of 2 from calculations within the $\Delta$-hole model. This
difference can only clarified by a measurement.
\par The agreement of the calculated total eta production cross section with the
experiment is good. The angular and energy differential cross sections show
a systematic deviation from the experiment for all nuclei and photon
energies. This can be corrected by a phenomenological adjustment of
the cross section for $\eta\, N$-scattering but it might as well be
due to an effect of the eta or nucleon potential which
is not accessible in our semi-classical calculation or to shortcomings in 
our description of $\eta$-rescattering.
\par An experimental measurement of exclusive cross sections, i. e. one-pion
production, two-pion production, nucleon knock-out etc., is imperative for
a better understanding of the photon-nucleus reaction and an explanation
of the disappearance of the $D_{13}$-resonance in the total photonuclear
absorption cross section. Such measurements would help to separate
possible in-medium modifications of the elementary $(\gamma,N)$ reactions
from final state interactions of the produced particles.
\par Nucleon knock-out reactions are of particular interest with respect
to multi-body absorption mechanisms. A comparison between our model
and experiments performed in Mainz \cite{cross95} will be available soon.

\newpage
\clearpage
\clearpage
\clearpage
\clearpage
\clearpage
\clearpage
\clearpage
\clearpage
\clearpage
\clearpage
\clearpage
\clearpage
\end{document}